\documentclass[aps,prb,twocolumn,letter]{revtex4}
\usepackage{epsfig}
\usepackage{amsmath}      
\usepackage{amssymb}      
\usepackage{float}        
\usepackage{graphicx}     
\usepackage{epsfig}
\begin{document}
\title{{\it Ab initio} tensile tests of Al bulk crystals and grain
  boundaries: universality of mechanical behaviour}  
\author{Rebecca Janisch}\email{rebecca.janisch@rub.de}
\author{Naveed Ahmed} 
\author{Alexander Hartmaier}
\affiliation{Interdisciplinary Centre for Advanced Materials
  Simulation, Ruhr-University Bochum, 44780 Bochum, Germany} 
\date{\today}
\begin{abstract}
  We have performed {\it ab initio} tensile tests of bulk Al along
  different tensile axes, as well as perpendicular to different grain
  boundaries to determine mechanical properties such as interface
  energy, work of separation and theoretical strength. We show that
  all the different investigated geometries exhibit
  energy-displacement curves that can be brought into coincidence in
  the spirit of the well known universal binding energy relationship
  curve. This simplifies significantly the calculation of {\it ab
    initio} tensile strengths for the whole parameter space of grain
  boundaries.
\end{abstract}
\maketitle
\section{Introduction}
The prediction of the mechanical properties of polycrystals requires
knowledge about the mechanical properties of all interfaces,
i.e.~grain boundaries, in their microstructure. So far, most mesoscale
models of microstructure-property relationships as used in continuum
simulations of deformation and fracture make rather simple assumptions
for the variation in grain boundary properties with the boundary
geometry\cite{Wei04,Kraft08}. With {\it ab initio} electronic structure
calculations it is possible to determine the cohesive energy, elastic
modulus, sliding barrier, and theoretical strength of interfaces
accurately and quantitatively. The considerable computational effort,
in comparison e.g.~to atomistic simulations employing empirical
potentials, is easily feasible with modern computers if the
investigations are restricted to grain boundary structures based on
coincidence-site lattices. The {\it ab initio} approach is desirable to
avoid problems of transferability of empirical potentials, usually
fitted to reproduce equilibrium properties, to non-equilibrium
processes such as failure. It is indispensable whenever a phenomenon
is controlled by the electronic structure. This is the case in
systems with directional bonds, or when studying the influence of
chemical composition, and alloying effects. Nevertheless, sampling the
five parameter space given by the geometric degrees of freedom of the
grain boundary (rotation axis and angle, and the grain boundary
normal\cite{Sutton95}) by {\it ab initio} calculations remains a
challenge. Models exist that relate the energies of certain subsets of
this space to the grain boundary geometry. Most well known is the
dislocation model for small angle tilt grain boundaries based on the
picture of Read and Shockley, which can be extended empirically to
large-angle tilt grain boundaries \cite{Wolf89}. If the energy of a
grain boundary can be related to its geometric parameters via such
models, the description of the energy hypersurface is significantly
simplified. However, so far it seems that there is no such correlation
which would be valid in the complete parameter space
\cite{Sutton87,Olmsted09}. The complexity of the problem is increased
by the fact that for use in mesoscale models we are not only looking
for a function that describes the grain boundary energy as function of
misorientation, but also for its first and second derivatives with
respect to a displacement from the equilibrium volume, i.e.~the
maximum stress and the elastic modulus. Nevertheless, the situation is
not hopeless, as the analytic function of this energy-displacement curve
is the same for any grain boundary geometry, if obtained under the
same loading conditions. The demonstration of this fact, a universal
binding behaviour in the spirit of Rose's universal binding energy
relationship (UBER) \cite{Rose83}, is subject matter of this paper.

In section \ref{sec:technical} we describe our computational
procedure, including an explanation of grain boundary nomenclature and
details on different ways to perform ``{\it ab initio}'' tensile tests. In
section \ref{sec:uber} we review the UBER and its implications. In the
results' section, sec.~\ref{sec:results}, we give details about the
grain boundary structures after a full optimization of the microscopic
degrees of freedom (\ref{sec:structures}) and the corresponding
energies (\ref{sec:energies}). The results of the tensile tests are
presented in section \ref{sec:displacements}. The universal elastic
behaviour under tensile load of all systems investigated is
demonstrated in section \ref{sec:scaling}, and the relationship
between energies and strength is discussed in section
\ref{sec:strength}.  We summarize our insights in section
\ref{sec:summary}.
\section{Technical details\label{sec:technical}}
For a given orientation of the tensile axis, we constructed supercells
for bulk, surface, and grain boundary calculations of the same size and
shape. In other words, starting from a bulk supercell containing $N$
atomic layers, half of the planes were replaced by vacuum to create a
surface slab, or half of the planes were replaced by the same number,
but with a misorientation, to create a grain boundary structure. 

In detail, supercells for tensile tests were constructed for Al bulk,
such that the $z$-axis is oriented along the [111], [112], [113], and
[114] direction, i.e.~defining (111), (112), (113) and (114) as the
cleavage plane. In addition we constructed special grain boundaries
containing these planes as grain boundary planes. These were the
$\Sigma$3 (111) [111] 60$^{\circ}$ twist grain boundary, the $\Sigma$3
(1$\bar{\mathrm{1}}$2) [110] 109$^{\circ}$ symmetrical tilt grain
boundary (STGB), the $\Sigma$11 (1$\bar{\mathrm{1}}$3) [110]
129$^{\circ}$ STGB, and the $\Sigma$9 (1$\bar{\mathrm{1}}$4) [110]
141$^{\circ}$ STGB. In this nomenclature the use of $\Sigma$ indicates
that for the chosen misorientation a periodic superstructure can be
found, the so-called coincidence site lattice (CSL). Therefore these
grain boundaries are also called special grain boundaries. The value
of $\Sigma$ is the volume of a unit of this CSL divided by the volume
of the cubic Al unit cell, i.e.~it is a measure for the periodicity of
the grain boundary. The grain boundary plane is given in round
brackets, the direction of the axis of misorientation in rectangular
ones. All tilt grain boundaries considered here are symmetrical, which
means that the grain boundary plane divides the misorientation angle
in two equal parts. In other words, it represents a mirror plane. Note
that the (111) [111] 60$^{\circ}$ twist grain boundary can also be
expressed as (111) [110] 70.5$^{\circ}$ tilt grain boundary, but we
prefer to refer to it as twist grain boundary to emphasize its
close-packed atomic structure.

To calculate the total energy of the above mentioned supercells we
performed density-functional-theory (DFT) calculations of total energy and
electronic structure employing the ABINIT open source code
\cite{abinit}. The exchange correlation effects were treated in the
local density approximation and electron-ion interactions were
modeled via a Trouller-Martins type norm-conserving pseudopotential
for Al. Convergency with respect to $k$-point density, plane wave
energy cut-off, and system size was tested for the (111) surface
energy. The surface energy is given by
\begin{eqnarray}
\gamma_{\mathrm{FS}} = \frac{2\cdot E_{\mathrm{tot}}^{\mathrm{FS}}-
  E_{\mathrm{tot}}^{\mathrm{bulk}}}{4A} \quad
. \label{eq:surface_energy}  
\end{eqnarray}
where $E_{\mathrm{tot}}^{\mathrm{FS}}$ is the total energy of the
surface slab and $E_{\mathrm{tot}}^{\mathrm{bulk}}$ the
energy of the bulk supercell containing twice the number of
atomic layers. $A$ is the surface area. The plane wave
cut-off was varied between 12 and 20 Ha in steps of 2 Ha. To test
convergency with respect to cell size, we chose bulk (surface)
supercells containing 18 (9), 24 (12), or 30 (15) (111) planes. The
$k$-point meshes employed were of the Monkhorst-Pack type, using
2x2x1, 4x4x1, 8x8x2, and 12x12x2 $k$-points. With a plane wave cut-off
of 16 Ha, the 8x8x2 Monkhorst-Pack mesh, and a minimum surface slab
thickness of nine atomic layers, the surface energy was converged
within an accuracy of 1$\cdot$10$^{-4}$ Ha/atom ($\approx$
2.5$\cdot$10$^{-3}$eV/atom).

For orientations different from $z$ parallel to [111] it was ensured
that the distance between the interfaces was at least as large as in
the [111] cells. The supercell shapes were made commensurate with that
of the [111] cell, which also enables the use of a commensurate
$k$-point mesh.

To obtain accurate interface energies all microscopic degrees of freedom
were optimized and the positions of the atoms were relaxed until the
remaining forces were smaller than 5$\cdot$10$^{-5}$
Ha/a.u.~($\approx$3$\cdot$10$^{-3}$ eV/\AA\ ). Afterwards, the excess
interplanar spacing of the grain boundaries can be calculated as half
the difference (due to the periodic boundary conditions) between the
relaxed ($D_{\mathrm 0}$) and initial ($D_{hkl}$) supercell length
perpendicular to the interface,
\begin{eqnarray}
\label{eq:expansion}
d_{\mathrm{0}} = \frac{D_{\mathrm{0}} - D_{hkl}}{2}
\end{eqnarray}
The grain boundary energy is given by
\begin{eqnarray}
\gamma_{\mathrm{GB}} =\frac{E_{\mathrm{tot}}^{\mathrm{GB}}- E_{\mathrm{tot}}^{\mathrm{bulk}}}{2A}\quad,
\end{eqnarray}
where $E_{\mathrm{tot}}^{\mathrm{GB}}$ is the total energy of the
grain boundary supercell and $E_{\mathrm{tot}}^{\mathrm{bulk}}$ the
energy of the bulk supercell containing the same number of
crystallographic layers in the corresponding orientation. $A$ is the
interface area. From the total energy of the surface slabs, also the
work of separation of a grain boundary was calculated according to
\begin{eqnarray}
W_{\mathrm{sep}}=\frac{2\cdot E_{\mathrm{tot}}^{\mathrm{FS}}- E_{tot}^{\mathrm{GB}}}{2A}\quad,
\end{eqnarray}
Note that the work of separation for the perfect bulk corresponds to
twice the surface energy as defined in eq.~(\ref{eq:surface_energy}).

After a fit of the energy-displacement curves, the tensile strength
can be calculated as the slope in the inflection point:
\begin{eqnarray}
\sigma_{\mathrm{th}} = \left. \frac{dE}{d\Delta}\right|_{E^{''}(\Delta)=0}
\label{eq:sigma}
\end{eqnarray}
where $\Delta$ is the displacement from the equilibrium interplanar
distance. Initially, the tensile tests were performed in three
different ways:
(a) by a simple scaling of the supercell dimensions along the
  tensile axis, while leaving the internal coordinates fixed. Atomic
  relaxations then lead to a homogeneous strain distribution in bulk
  supercells, and a characteristic distribution of strain in any cell
  containing a defect (e.g.~a grain boundary).
(b) By performing rigid grain shifts (rgs). Here the spacing between
  two blocks of atoms is increased only at a defined cleavage plane
  (due to the use of periodic boundary conditions at two defined
  cleavage planes per supercell). Within the blocks, the interplanar
  distance corresponds to the equilibrium bulk value. This way we can
  model ideally brittle cleavage under loading mode I.
(c) By doing the latter and relaxing the atomic positions
  at each shift, while keeping the total elongation of the supercell
  fixed. Again, this corresponds to a mode I cleavage process, but now
  elastic energy is released due to atomic relaxations.
\section{Universal behavior and scaling lengths\label{sec:uber}}
Rose et al.\cite{Rose83}~postulated and demonstrated that the binding energies $E_b$ of metals
have a universal form of the kind 
\begin{eqnarray}
E_{\mathrm{b}}(d) = |E_{\mathrm{b}}^{\mathrm{e}}|g(a)\quad ,
\label{eq:uber}
\end{eqnarray}
where $d$ is the interatomic distance, or, in the case of interface
energies the interplanar spacing. $|E_{\mathrm{b}}^{\mathrm{e}}|$ is
the binding energy at equilibrium volume ($=-W_{{\mathrm{sep}}}$) and
$a$ is the rescaled displacement
\begin{eqnarray}
a = \frac{\Delta}{l}\quad .
\end{eqnarray}
The characteristic length scale $l$ depends on
the curvature of the energy-volume curve at the minimum, i.e., on the
elastic modulus as follows,
\begin{eqnarray}
\label{eq:l}
l = \sqrt{\frac{|E_{\mathrm{b}}^{\mathrm{e}}|}{E_{\mathrm{b}}^{''}(d_{\mathrm{0}})}}\quad .
\end{eqnarray}
If rescaled in this manner, all energy-volume curves coincide,
i.e.~they have the same functional form $g(a)$. This phenomenon was
observed for adhesion \cite{Ferrante79} and cohesion \cite{Rose84} of
metals, as well as chemisorption on metal surfaces \cite{Smith82}.
More recently, Hayes et al.~have shown that even the
energy-displacement curves of non metallic systems (Al$_2$O$_3$ and
Si) can be brought into coincidence with those of metals (Al)
\cite{Hayes04}. This universal behavior of chemically different
systems means that we can determine the cohesive behavior
(i.e.~theoretical strength and critical displacement) of any material
from three parameters, $E_{\mathrm{b}}^{\mathrm{e}}$, $d_{\mathrm{0}}$, and $E_{\mathrm{b}}^{''}(d_{\mathrm{0}})$, once the
functional form $g(a)$ is known.\\
To describe hydrostatic volume expansion / compression of simple
metals, Rose determined $g(a)$ to be
\begin{eqnarray}
g(a) = -(1+a+0.05a^3)e^{-a} \quad .
\end{eqnarray}
Hayes et al.~used an asymptotic approximation, i.e.~a simple quadratic
function introduced by Nguyen et al.\cite{Nguyen02} that scales with the
system size to fit results of uniaxial computational tensile tests. Originally
applied to the results of homogeneous strain, the approach was generalized by
Hayes et al.~to fit the results of tensile tests in the form of rgs + atomic
relaxation, thus taking into account surface relaxations\cite{Hayes04}. The simple function
and the scalability of the results make this approach attractive for the
calculation of traction-separation laws used in continuum models. However, if
the theoretical strength shall be calculated independently of the system size,
a function displaying an inflection point is to be preferred.  To fit the
results of our tensile tests in the form of rigid grain shifts, we used
\begin{eqnarray}
g(a) = -(1+a)e^{-a} \quad ,
\label{eq:g_tensile}
\end{eqnarray}
a function also used by Rose, to represent the results of displacing
metal-metal interfaces of different metals.
\section{Results\label{sec:results}}
\subsection{Grain boundary structures\label{sec:structures}}
After construction according to the macroscopic parameters the
microscopic degrees of freedom of the grain boundaries were optimized
by performing rigid grain shifts perpendicular as well as parallel to
the interface, followed by relaxation of the atomic positions. All
grain boundaries exhibit excess volume at the interface, calculated
according to equation (\ref{eq:expansion}).  These expansions
$d_{\mathrm{0}}$ are localized in the vicinity of the grain boundary,
but not necessarily confined only to the first crystallographic plane
at the interface. The expansions are summarized in table
\ref{tab:excess}. The smallest expansion is observed for the (111)
twist grain boundary, which is the most dense cleavage plane after the
(111) bulk plane.
\begin{table}
\centering
\begin{ruledtabular}
\begin{tabular}{lrrr}
                                           &$d_0$ [\AA]& $l_b$ [\AA]& $l_r$[\AA]\\ \hline
(111) bulk                                 & -         & 0.561      & 1.652     \\    
(112) bulk                                 & -         & 0.653      & 1.599     \\
(113) bulk                                 & -         & 0.655      & 1.695     \\
(114) bulk                                 & -         & 0.665      & 1.657     \\ \hline
$\Sigma 3 (111)  [111] 60^{\circ}$ twist GB & 0.014     & 0.615      & 1.745     \\
$\Sigma 3 (1\bar{1}2)[110]109^{\circ}$ STGB & 0.389     & 0.621      & 1.736     \\
$\Sigma 11(1\bar{1}3)[110]129^{\circ}$ STGB & 0.311     & 0.620      & 1.508     \\
$\Sigma 9 (1\bar{1}4)[110] 141^{\circ}$ STGB & 0.185    & 0.605      & 1.514    \\
\end{tabular}
\end{ruledtabular}
\caption{\label{tab:excess} Excess volume at the different grain
  boundaries, represented by $d_0$, and scaling lengths for brittle
  and relaxed cleavage, $l_b$ and $l_r$, for bulk and grain boundary
  planes (explanation in the text).}   
\end{table}
The stable translation state of the $\Sigma$11 STGB parallel to the
interface is the initial, mirror-symmetric one. In the case of the
$\Sigma$3 STGB a shift of 0.5 times the interplanar spacing along the
tilt axis [110] breaks this mirror-symmetry and produces a structure
about 150 mJ/m$^2$ lower in energy than the mirror-symmetric one. The
$\Sigma$9 STGB initially contained two atomic columns at unphysically
small distance at the grain boundary. Relaxing the structure (rgs +
atomic relaxations) led to major re-arrangements of the atomic
positions at the interface. Removing instead one atomic column at the
grain boundary and relaxing the structure (rgs + atomic relaxation)
led to an interface energy which is lower by 0.067 Ha/atom ($\approx$
1.8 eV/atom). The resulting structure at the grain boundary is in
excellent agreement with experimental observations \cite{Muller99}.
\subsection{Interface energies and work of separation\label{sec:energies}}
The resulting interface energies, i.e.~surface and grain boundary
energies, and the work of separation for the systems investigated are
shown in table \ref{tab:energies}.  
\begin{table}
\begin{ruledtabular}
\begin{tabular}{lrrr} 
                  &$\gamma$ & $W_{\mathrm{sep}}$ & $\sigma_{\mathrm{th}}$\\
                  &[J/m$^2$]& [J/m$^2$] & [GPa]\\          
\hline
(111) bulk                                 & 0.987 & 1.973 & 12.6 \\
(112) bulk                                 & 1.109 & 2.218 & 12.8 \\
(113) bulk                                 & 1.093 & 2.187 & 12.7 \\
(114) bulk                                 & 1.145 & 2.290 & 13.2 \\\hline
$\Sigma 3 (111)  [111] 60^{\circ}$ twist GB & 0.048 & 1.924 & 11.5 \\
$\Sigma 3 (11\bar{2})[110]109^{\circ}$ STGB & 0.393 & 1.818 &  6.8 \\
$\Sigma 11(1\bar{1}3)[110]129^{\circ}$ STGB & 0.171 & 2.007 & 12.0 \\
$\Sigma 9(1\bar{1}4)[110] 141^{\circ}$ STGB & 0.486 & 1.779 & 12.1  \\
\end{tabular}
\end{ruledtabular}
\caption{\label{tab:energies} Interface energy, work
  of separation, and theoretical strength of Al bulk and grain boundaries.}  
\end{table}
The grain boundary energies follow the known trend for the energies of
symmetric [110] tilt grain boundaries in fcc metals as function of the
misorientation angle, with energy cusps representing the $\Sigma$3
twist and the $\Sigma$11 STGB
\cite{Wolf89,Sutton95,Saylor04}. Results of previous DFT
investigations were found for the $\Sigma$3 twist \cite{Lymperakis09}
and the $\Sigma$11 STGB \cite{Wright94} and agree well with our
results. The work of separation for the grain boundaries does not
follow the same trend, as it also depends on the corresponding surface
energy. The lowest surface energy in Al is that of the close packed
(111) plane. Thus, $W_{\mathrm{sep}}$ for the stable twin is lower
that that of the $\Sigma$11 STGB, corresponding to the second-most
favorable [110] grain boundary orientation in fcc metals.
\subsection{Energy-displacement curves\label{sec:displacements}}
\begin{figure}
\centering
\includegraphics[width=7.8cm]{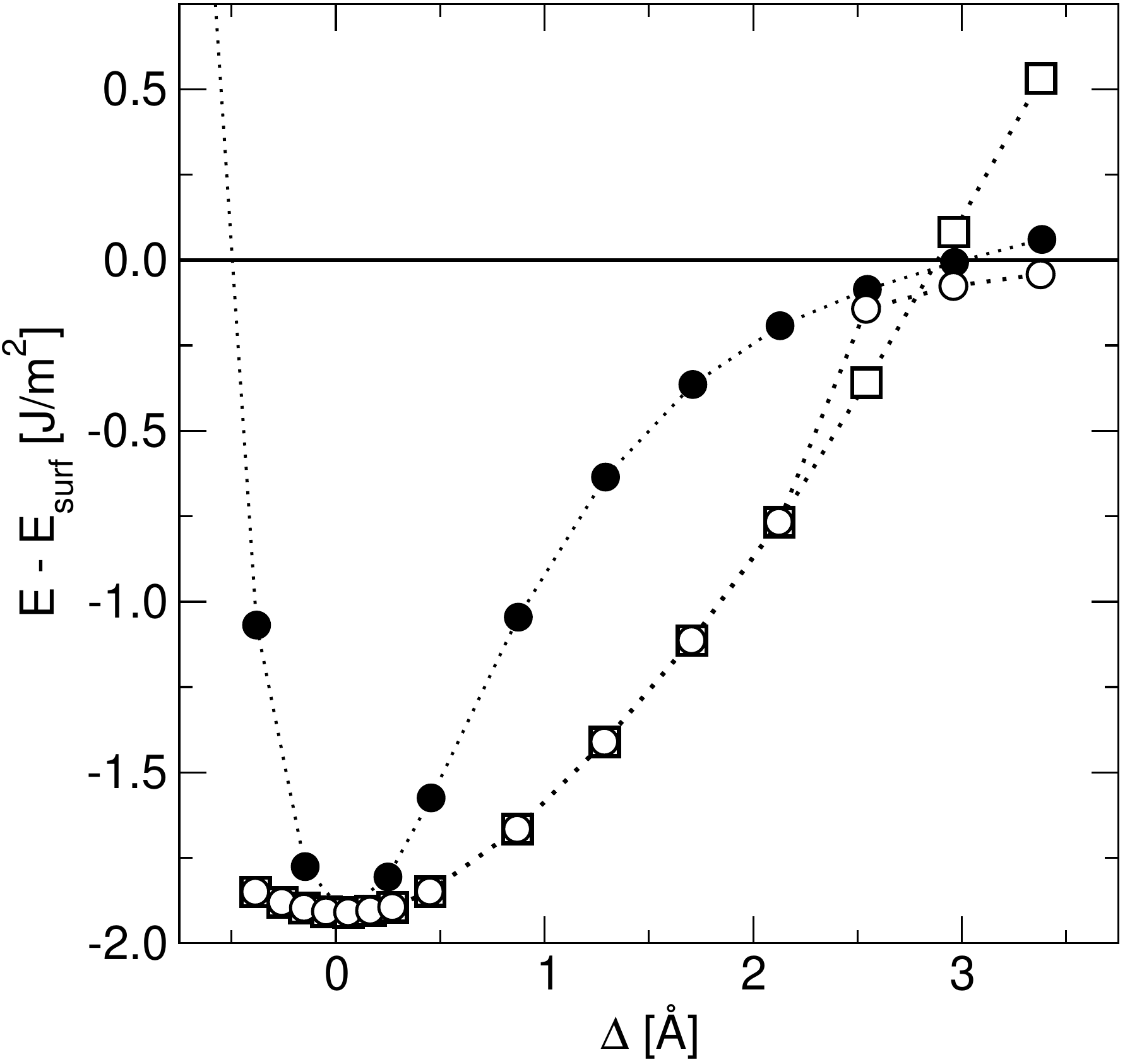}
\caption{\label{fig:111-tensile-energies} Results of {\it ab initio} tensile
  tests along the [111] direction in bulk Al. Black circles represent
  the results of rgs, open circles that of rgs followed by relaxation
  of the atomic positions, and squares that of a homogeneous straining of
  the supercell.}
\end{figure}
Starting from the optimized structures, tensile tests were performed
as explained in section \ref{sec:technical}.  Figure
\ref{fig:111-tensile-energies} shows the results for bulk Al stretched
and compressed along the [111] direction. Rigid grain shifts without
any atomic relaxation produce a rather steep curve in a shape that can
be fitted well by equation (\ref{eq:g_tensile}). The energy
asymptotically approaches the unrelaxed (111) surface energy, which is
slightly higher than the relaxed one taken as the reference energy
level in fig.~\ref{fig:111-tensile-energies}. After relaxing the
atomic structure at each shift, the atomic positions and the
corresponding energies coincide with the results obtained by
stretching the supercell homogeneously. A deviation is observed above
a relative displacement of the two bulk slabs of 2\AA. In the case of
the rgs, the atoms relax to a configuration where the strain is
localized at the defined cleavage plane, whereas in case of the
homogeneous strain the crystallographic planes have the same
distance. A continuation of the homogeneous tensile test would lead to
$N$ free standing crystallographic planes, if $N$ is the number of layers
in the supercell.

\begin{figure}
\centering
\includegraphics[width=7.8cm]{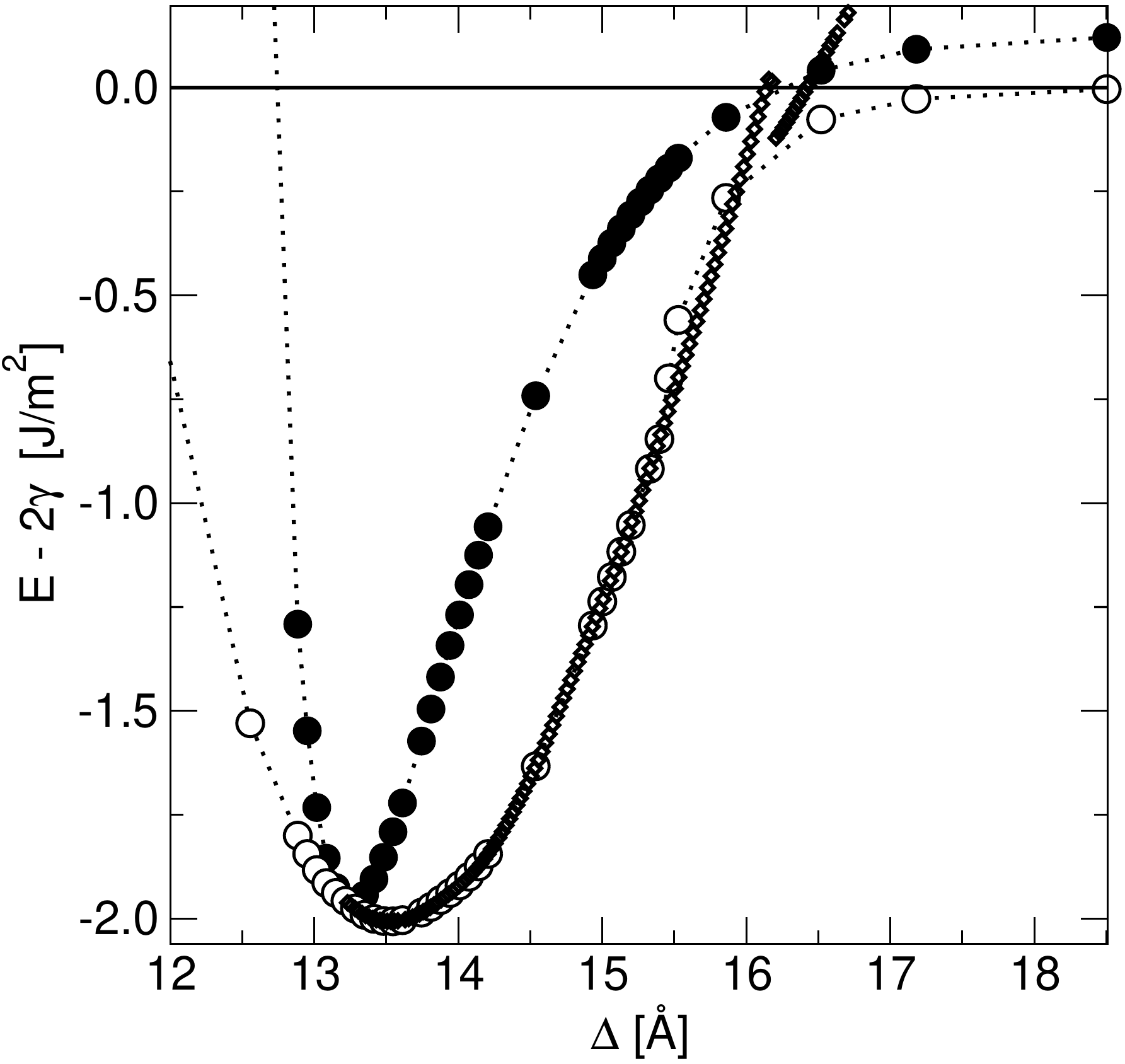}
\caption{\label{fig:S11-tensile-energies} Results of {\it ab initio} tensile
  tests at the $\Sigma$11 STGB. Black
circles represent the results of rgs calculations
without atomic relaxations, and open circles mark the energies after
atomic relaxation. Diamonds are the energies obtained via a
``drag'' calculation (see text). The reference level is the energy of the
relaxed (113) surface.} 
\end{figure}

If the tensile test is performed with a supercell containing a grain
boundary, the strain is expected to localize at this defect also in
the case of a homogeneous elongation of the supercell. This is shown
in figure \ref{fig:S11-tensile-energies} for the $\Sigma 11 (113)
[110] 129^{\circ}$ symmetrical tilt grain boundary. Again, the black
circles represent the results of the rigid grain shift calculations
without atomic relaxations. The open circles mark the energies after
atomic relaxation. The diamonds indicate the energies obtained after
a successive scaling of the total length of the supercell,
i.e.~starting with the equilibrium structure the relaxed reduced
coordinates of the atoms are taken as input coordinates of the next
strain state. This ensures a continuous deformation path up to the
point where the energy crosses the reference level, the energy of the
relaxed (113) surface. At this point the energy drops and then starts
to increase again. An analysis of the atomic structure at this
discontinuity shows that the two grains have partially debonded and are
only connected via chains of atoms containing the initial coincidence
site. Due to this coincidence site a further elongation of the cell
again leads to a free standing crystallographic plane, in the center
between two (113) surfaces. Thus, while such a drag calculation is
interesting to investigate failure modes and critical displacements,
it is not suitable to calculate the work of separation or the tensile
strength. We can conclude that for the latter a defined point of
failure is needed in every loading scheme.
\subsection{Scaling lengths for different geometries\label{sec:scaling}}
\begin{figure}
\centering
\includegraphics[width=7.8cm]{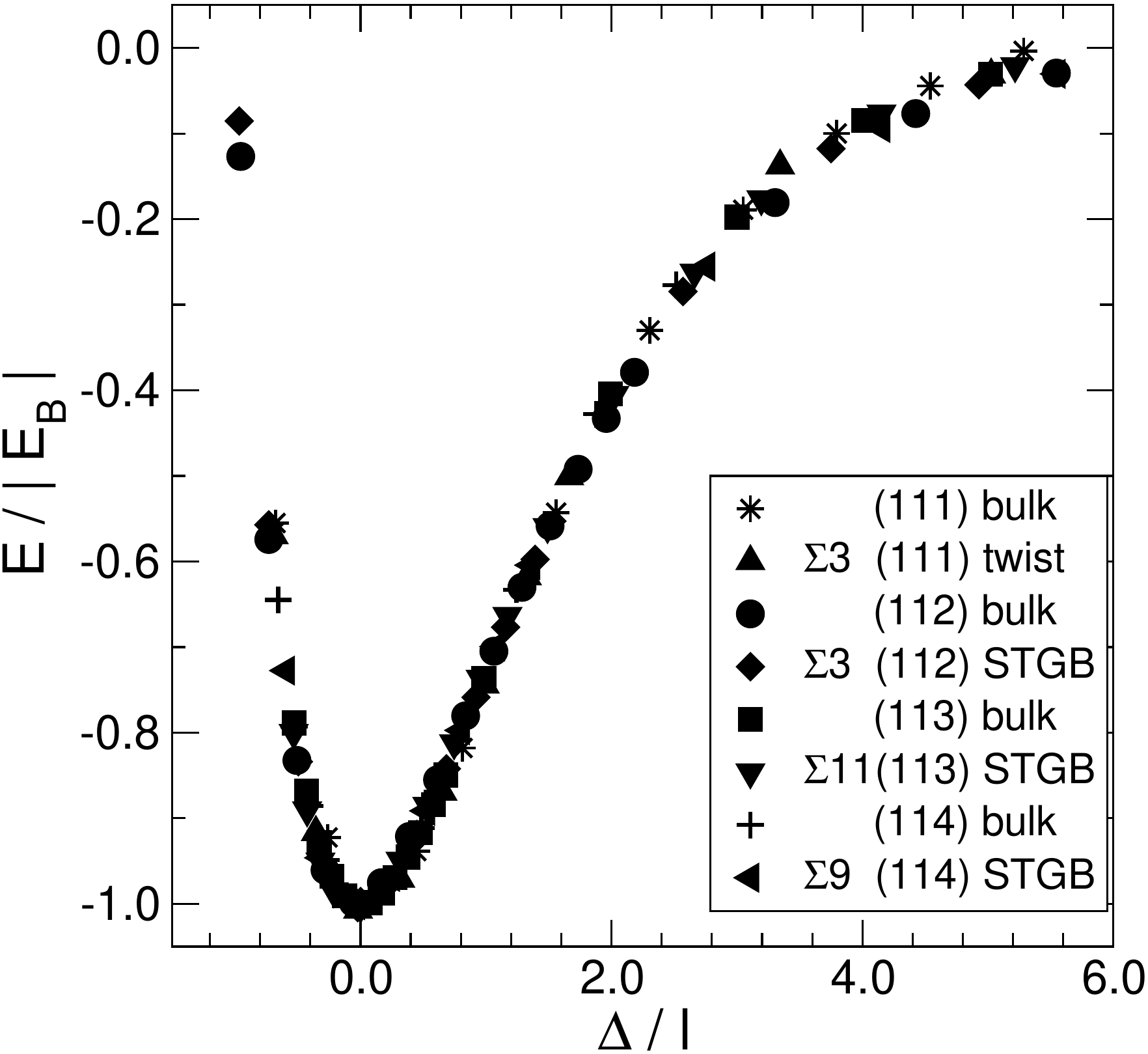}
\caption{\label{fig:uber} Rgs energy displacement relationships
  re-scaled according to eqs.~\ref{eq:uber} to \ref{eq:l}, using
  eq.~\ref{eq:g_tensile}. For details of the twist and STGB geometries see text.} 
\end{figure}

\begin{figure}
\centering
\includegraphics*[width=7.8cm]{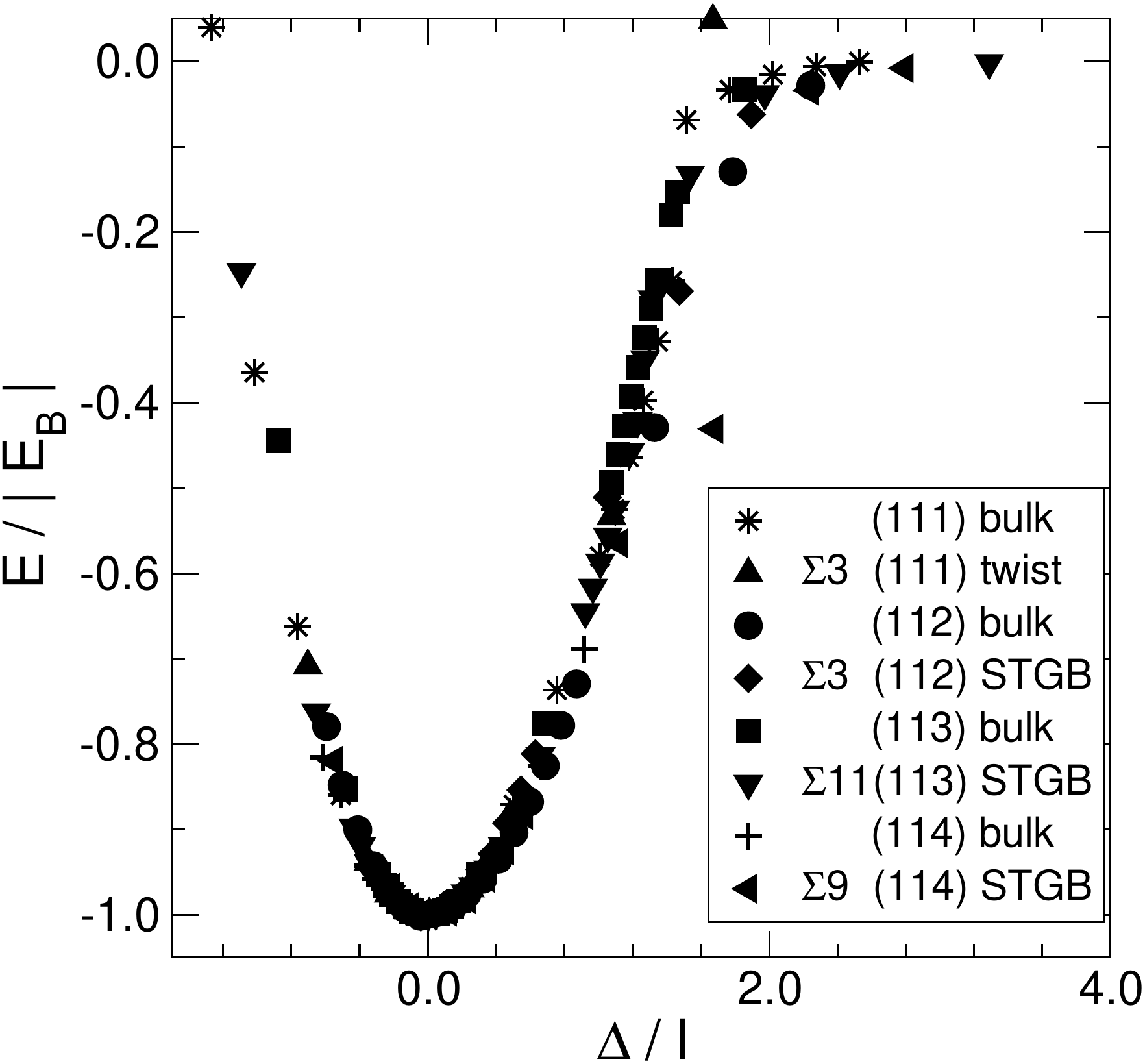}
\caption{\label{fig:uber_relaxed} Energy displacement relationships
  with relaxed atomic positions, re-scaled according to
  eqs.~\ref{eq:uber} to \ref{eq:l}, using the harmonic approximation
  for $g(a)$. For details of the twist and STGB geometries see text.}
\end{figure}

As mentioned in section \ref{sec:uber}, a universal behavior of
materials under strain has been shown for bulk metals and ceramics,
and for coherent metal-metal interfaces. In our investigation, we
stick to one metal, Al, but probe different orientations of the tensile
axis for bulk tensile tests, as well as different grain boundary
geometries, where the tensile axis is perpendicular to the
interface. As can be seen in figure \ref{fig:uber} the results of rgs
calculations scale perfectly. The parameters resulting from the fit
are given in table \ref{tab:energies}
($E_{\mathrm{b}}=-W_{\mathrm{sep}}$).
For the case of rgs followed by atomic relaxations, the function given
in equation (\ref{eq:g_tensile}) does not describe the results
well. The fit could be improved by using a polynomial including higher
order terms ($\geq$3). However, for the rescaling procedure the exact
function $g(a)$ actually does not have to be known. Instead we
determined the curvature at minimum energy by assuming a quadratic
function close to the minimum, i.e.~making a harmonic
approximation. The results of rgs calculations followed by atomic
relaxations were rescaled with the curvature thus obtained and with
the binding energy. The rescaled curves are shown in figure
\ref{fig:uber_relaxed}. Apart from a few outliers the agreement is
satisfactory. Deviations occur mainly at displacements where the
structure becomes unstable and where we can think of a crack forming.
Thus, whether investigating relaxed or unrelaxed cleavage of different
geometries, all we have to know about a system to describe its elastic
response upon tensile load are the binding energy in the specified
cleavage plane and the elastic constant (resp.~$E^{''}(\Delta)$) for
the tensile direction. Especially for the grain boundaries we can
conclude that once we have determined a fitting function $g(a)$ we can
use it for the complete parameter space spanned by the degrees of
freedom of the interfaces.

The scaling lengths calculated to make the different
energy-displacement curves coincide are given in table
\ref{tab:excess}.  For ideally brittle cleavage (rgs without atomic
relaxation), the results for bulk Al agree very well with the
empirical results of Rose et al.~\cite{Rose83}, 0.66\AA. This value
was obtained by using the experimental results of Simmons and
Wang\cite{Simmons71} for the elastic constant of Al along the [111]
direction and the surface energy of Tyson and
Miller\cite{Tyson77}. The latter represents an average over different
crystallographic planes, thus we expect our value to be more
exact. {\it Ab initio} calculations for bulk Al have been carried out by
Lazar et al.\cite{Lazar08}. For tensile tests along the [111]
direction they obtain $l_b$ = 0.54 \AA\ and $l_r$ = 2.4 \AA. While
there is excellent agreement between the scaling lengths for brittle
cleavage, our result for the relaxed cleavage, also shown in table
\ref{tab:excess}, is about 30\% lower. This is due to the fact that,
although the energy-displacement curve is fitted to a quadratic
function in both cases, the definition of $l_r$ is different.  Lazar
and Podloucky \cite{Lazar08} define $l_r$ as the critical opening, at
which the elastic energy $E(\Delta)$ equals the work of separation,
comparable to the discontinuity in our drag results, see figure
\ref{fig:111-tensile-energies}. This means their quadratic fit is
performed in the spirit of the asymptotic approximation of Nguyen and
Ortiz \cite{Nguyen02}, however it is independent of the system
size. In our case, where the quadratic fit is strictly confined to the
region around the energy minimum, $l_r$ has no such meaning, but is
simply a measure for the stiffness of the cleavage plane, as in the
work of Rose et al.
\subsection{Tensile strength \label{sec:strength}}
From the derivatives of the energy-displacement curves, the
theoretical tensile strength of the bulk phase along a given tensile axis
and of grain boundaries along an axis perpendicular to the grain
boundary plane, can be calculated according to equation
(\ref{eq:sigma}).  For direct comparison of bulk and grain boundary
properties, we restrict ourselves to the results of rgs calculations
without relaxation of the atomic positions. Thus, the strain is fully
localized between the cleavage planes in both cases. A relaxation
would lead to different strain distributions in the supercell, as
discussed in detail in section \ref{sec:technical}. The results are
summarized in table \ref{tab:energies}. The theoretical strength of a
grain boundary is generally lower than that of bulk lattice planes in
the corresponding orientation. Furthermore, in the bulk, the trend in
theoretical strength follows the one in the work of separation: the
higher $W_{\mathrm{sep}}$, the higher $\sigma_{\mathrm{th}}$. For the
grain boundaries, however, although the same overall trend is still
visible, the relationship is not so straight forward.
\section{Summary and Conclusions\label{sec:summary}}
We have performed {\it ab initio} tensile tests of bulk Al along different
tensile axes, as well as perpendicular to different grain
boundaries. It has been discussed that in order to simulate a physically
meaningful de-cohesion process, a plane of cleavage has to be defined
in all systems, also those containing defects as grain
boundaries. This was done by performing the tensile tests by means of
rigid grain shifts, followed or not by relaxation of the atomic
positions at each shift.
\begin{figure}
\centering
\includegraphics[width=7.8cm]{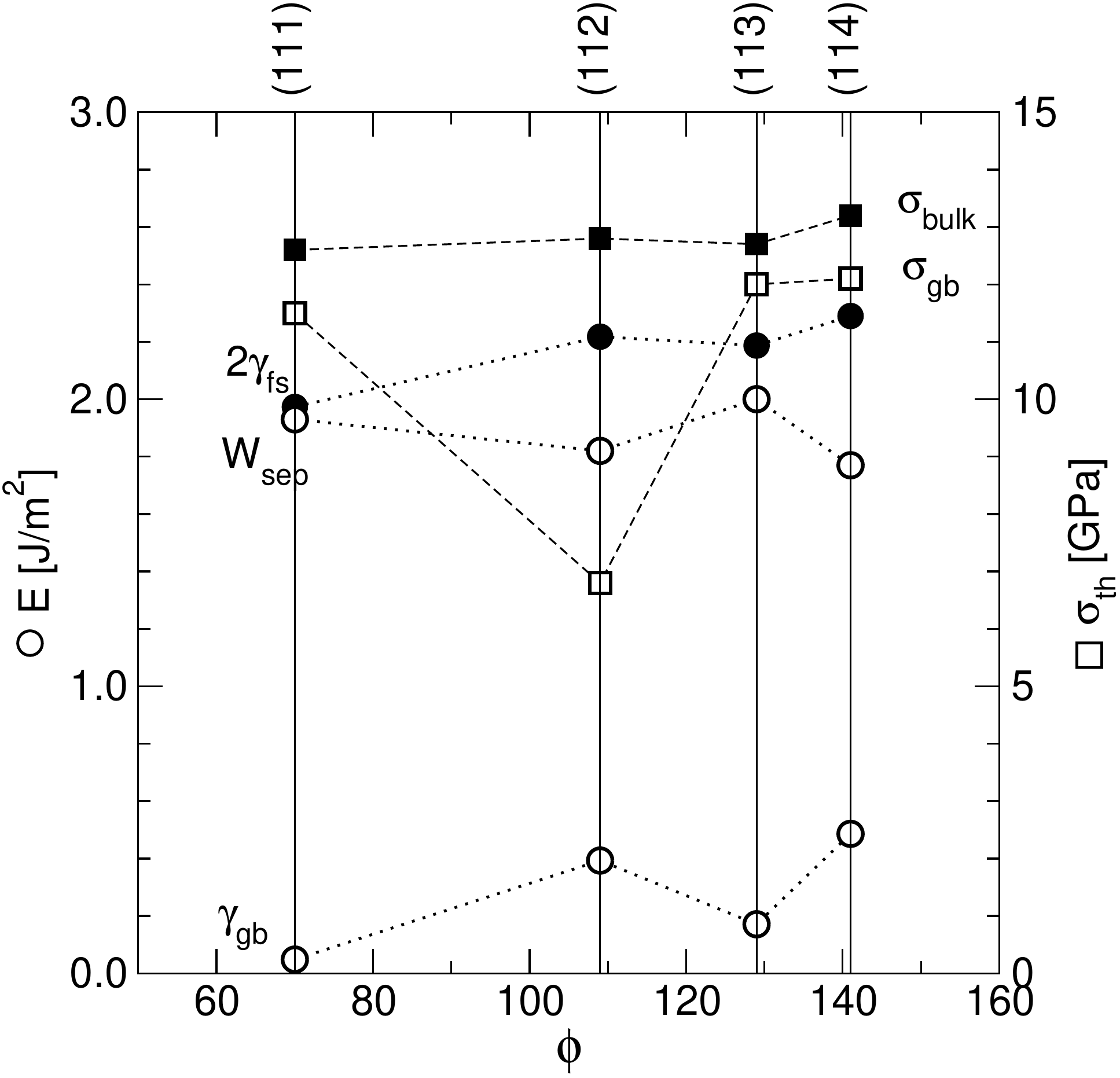}
\caption{\label{fig:properties} Mechanical properties of the different
  grain boundaries and bulk supercells as function of the
  misorientation from (110). Energies are represented by
  circles, and stresses by squares. Grain boundary properties are
  displayed in white, and bulk and surface properties in black. For more
  explanations see text.}
\end{figure}
The mechanical properties of the investigated systems are summarized
again in figure \ref{fig:properties}. The grain boundary energies
agree well with the empirical extension of the Read-Shockley picture,
showing cusps at the $\Sigma3$ (111) and the $\Sigma$11 (113)
orientations. The most stable defect judging from the interface energy
is the (111) twist grain boundary. However, this is not the grain
boundary with the highest strength, demonstrating that for defect
structures as grain boundaries, there is no simple relation between
energy and strength, and $\sigma_{\mathrm{th}}$  has to be
calculated explicitly.

Fortunately, it is not necessary to calculate full energy-displacement
curves for the complete parameter space of grain boundaries, because,
as we were able to show, these exhibit a universal behavior. Thus,
after having determined the analytical function $g(a)$ once, we only
need to determine the minimum energy structure, corresponding surface
energies, and the curvature at the minimum. By this simplification a
use of {\it ab initio} tensile strengths in continuum models of polycrystals
is getting within reach.
\acknowledgments 
The authors thank Christian Els\"asser for valuable comments on the
UBER. Furthermore the authors acknowledge financial support through
ThyssenKrupp AG, Bayer MaterialScience AG, Salzgitter Mannesmann
Forschung GmbH, Robert Bosch GmbH, Benteler Stahl/Rohr GmbH, Bayer
Technology Services GmbH and the state of North-Rhine Westphalia as
well as the European Commission in the framework of the European
Regional Development Fund (ERDF).
%
%
%
%

%
\end{document}